\documentclass[epj,nopacs]{svjour}
\usepackage{graphics}
\usepackage{times}
\def\ga{\mathrel{\mathchoice {\vcenter{\offinterlineskip\halign{\hfil
$\displaystyle##$\hfil\cr>\cr\sim\cr}}}
{\vcenter{\offinterlineskip\halign{\hfil$\textstyle##$\hfil\cr>\cr\sim\cr}}}
{\vcenter{\offinterlineskip\halign{\hfil$\scriptstyle##$\hfil\cr>\cr\sim\cr}}}
{\vcenter{\offinterlineskip\halign{\hfil$\scriptscriptstyle##$\hfil\cr>\cr
\sim\cr}}}}}
\newcommand{\NOTE}[1]{}
\newcommand{\FigRef}[1]{Fig.\,\ref{#1}}
\newcommand{\TabRef}[1]{Table\,\ref{#1}}
\newcommand{\FigDef}[1]{}
\newcommand{\TabDef}[1]{}
\newlength{\FigWidth}
\newcommand{\FigI}{
  \begin{figure}[ht]
   \includegraphics{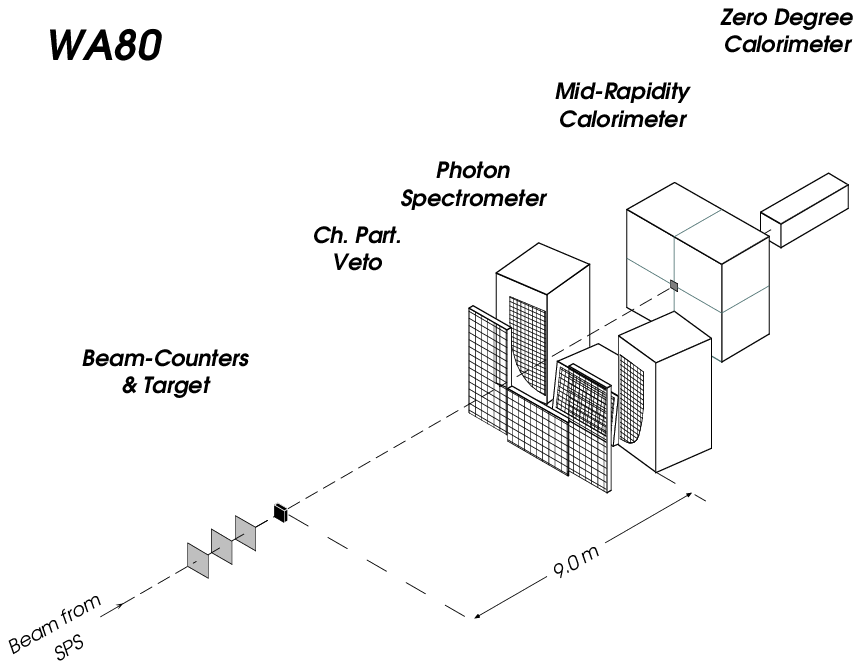}
   \caption{WA80 experimental setup}
   \protect\label{fig:wa80-setup}
  \end{figure}
  \FigDef{fig:wa80-setup}
}

\newcommand{\FigII}{
  \FigDef{fig:overlap}
  \begin{figure}[tbh]
	\includegraphics{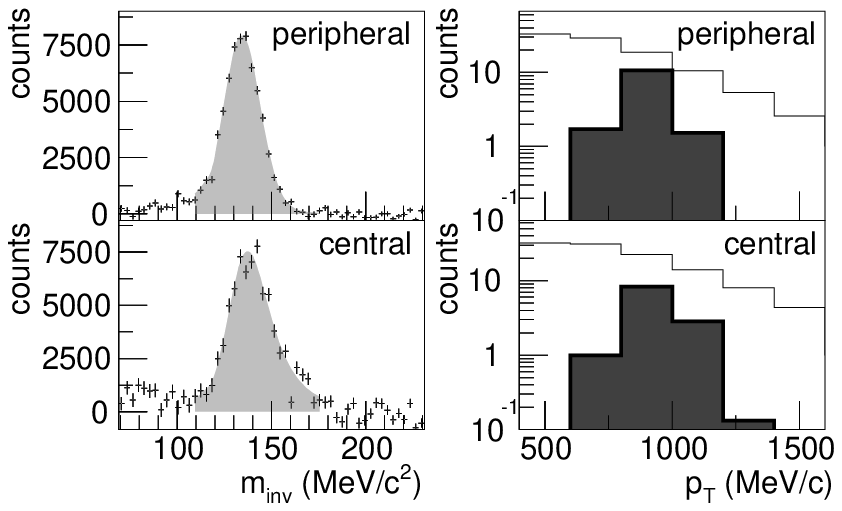}
	\caption{Influence of shower overlap on the neutral pions with 
	$0.8 \, \mathrm{GeV}/c \leq p_{T} \leq 1.0 \, \mathrm{GeV}/c$ for 
	peripheral and central collisions. The 
	plots on the left hand side show two-photon invariant mass spectra, 
	where the light grey area indicates the integration range for the 
	pion extraction. The dark grey histograms on the right hand side 
	show the $p_{T}$ distribution of pions simulated with 
	$0.8 \, \mathrm{GeV}/c \leq p_{T} \leq 1.0 \, \mathrm{GeV}/c$ after 
	overlap effects from other showers are included. For reference, the 
	relevant region of the $p_{T}$-spectra is also shown. See text for 
	further discussion.}
	\protect\label{fig:overlap}
  \end{figure}
}

\newcommand{\FigIII}{
  \FigDef{fig:acceptance}
  \begin{figure}[tbh]
	\includegraphics{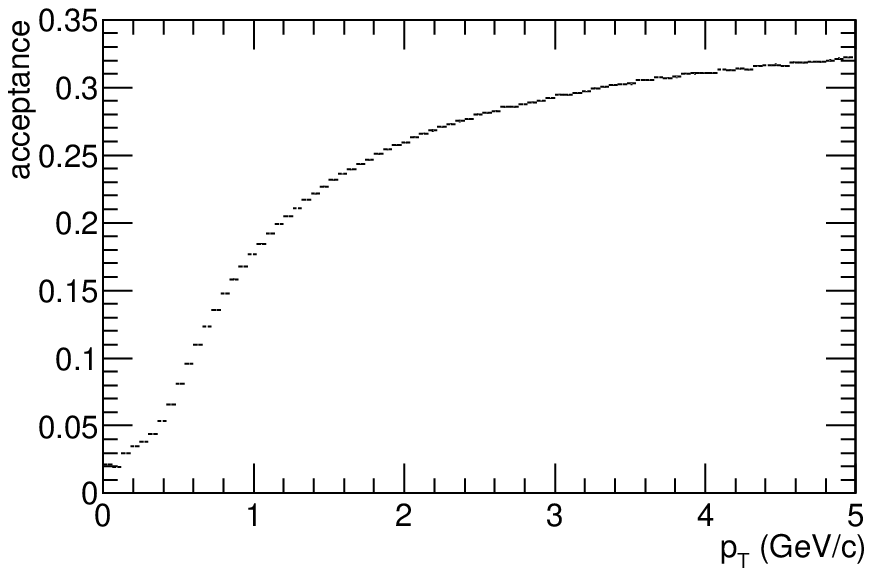}
	\caption{Geometrical acceptance for neutral pions as a function of 
	$p_{T}$.}
	\protect\label{fig:acceptance}
  \end{figure}
}

\newcommand{\FigIV}{
  \FigDef{fig:efficiency}
  \begin{figure}[b]
	\includegraphics{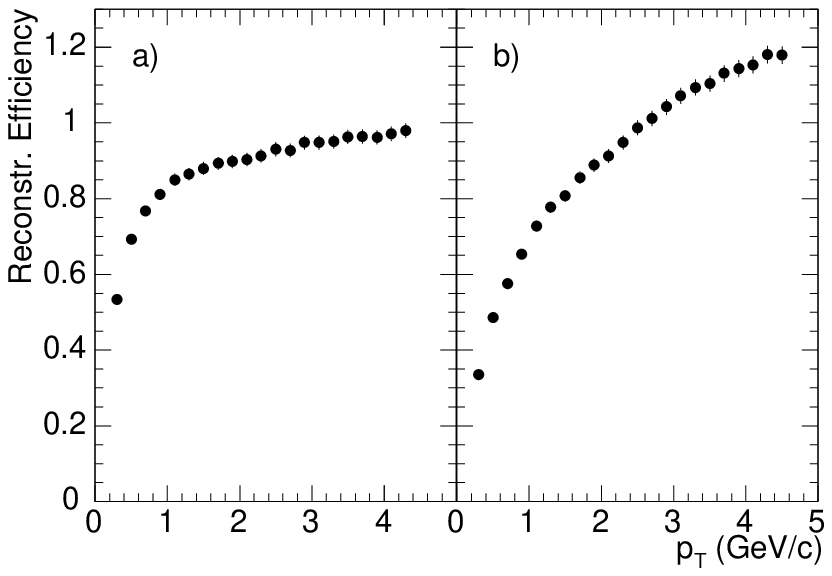}
	\caption{Neutral pion reconstruction efficiency for very peripheral 
		(a) and very central $^{32}$S~+~Au reactions (b).}
	\protect\label{fig:efficiency}
  \end{figure}
}

\newcommand{\FigV}{
  \FigDef{fig:fehler}
  \begin{figure}[h]
	\includegraphics{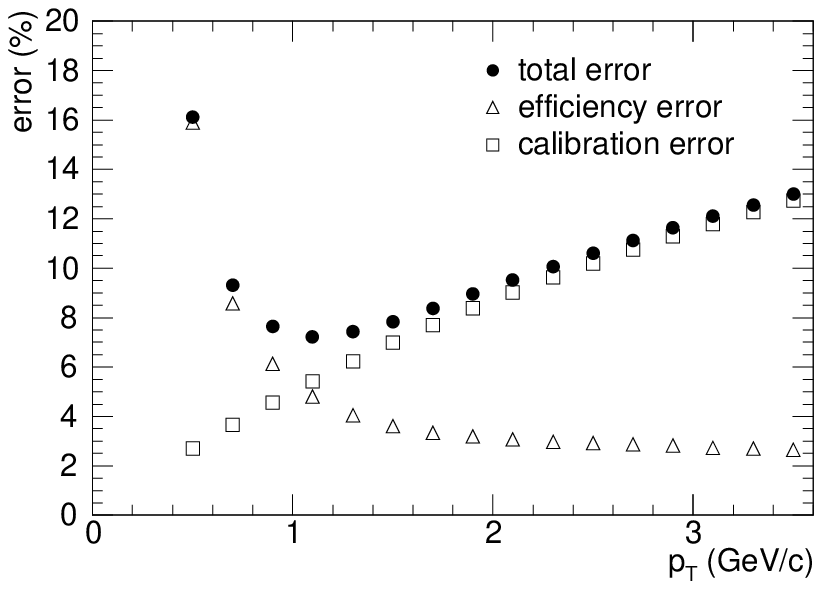}
	\caption{
		Systematic errors of the extraction of the $\pi^0$ yields
		for central S\,+\,Au collisions as a function of $p_T$.
	}
	\protect\label{fig:fehler}
  \end{figure}
}

\newcommand{\FigVI}{
  \begin{figure*}[tb]
    \includegraphics{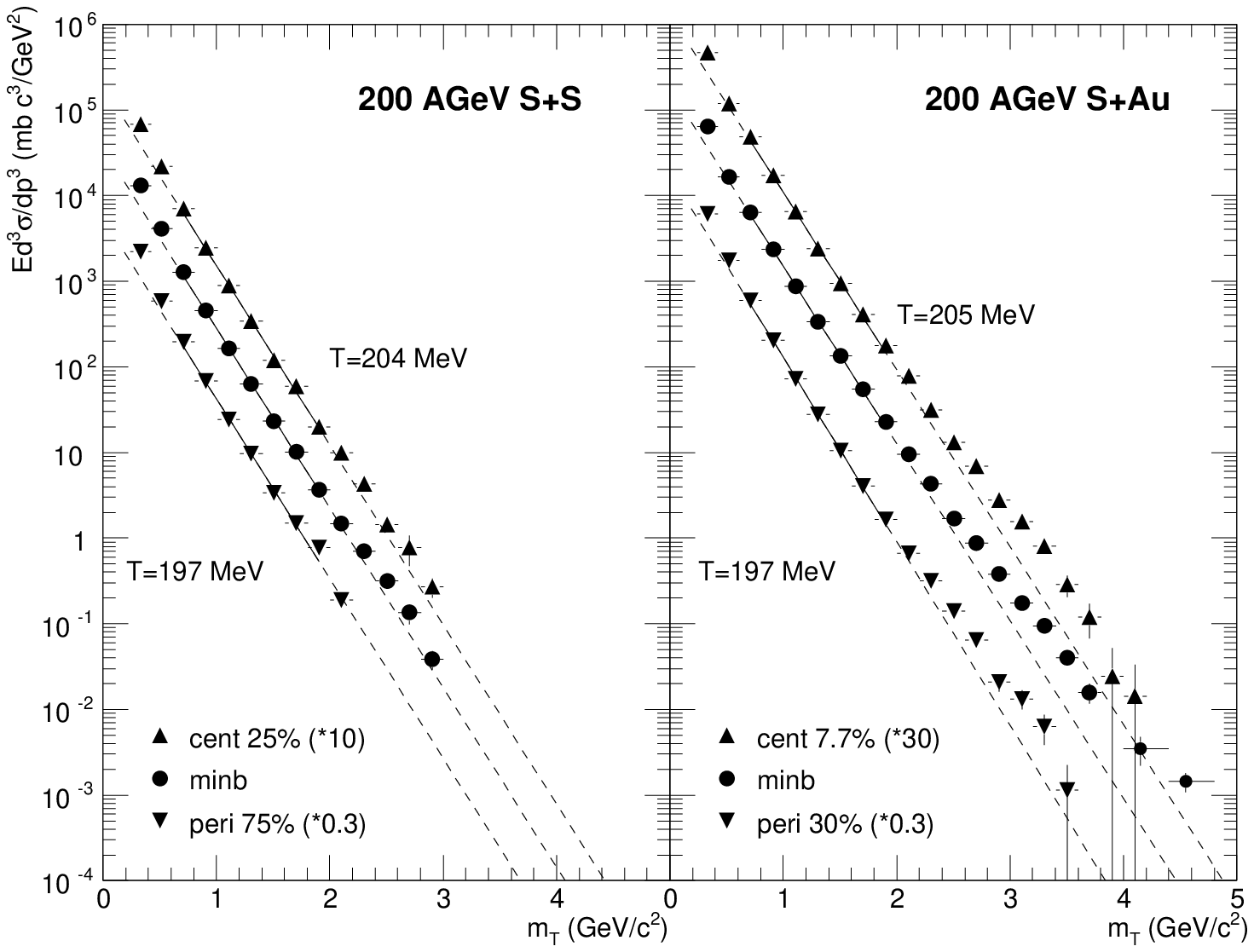}
    \caption{Invariant cross sections of $\pi^0$ mesons from 
	reactions of S\,+\,S (left) and S\,+\,Au (right) at
	200\,$A\cdot$GeV measured in the rapidity range $2.1 \leq
	y \leq 2.9$. The events are selected for centrality with the 
	percentage of the minimum bias cross section as indicated.  The 
	lines drawn with the data are exponentials fitted to the region 
	0.8 GeV/$c^{2}$ $\le m_T \le$ 2.0 GeV/$c^{2}$ with slope parameters as
	indicated.}
    \protect\label{fig:pt-ss-sau}
  \end{figure*}
\FigDef{fig:pt-ss-sau}
}

\newcommand{\FigVII}{
  \FigDef{fig:local-slopes}
  \begin{figure}[b]
	\includegraphics{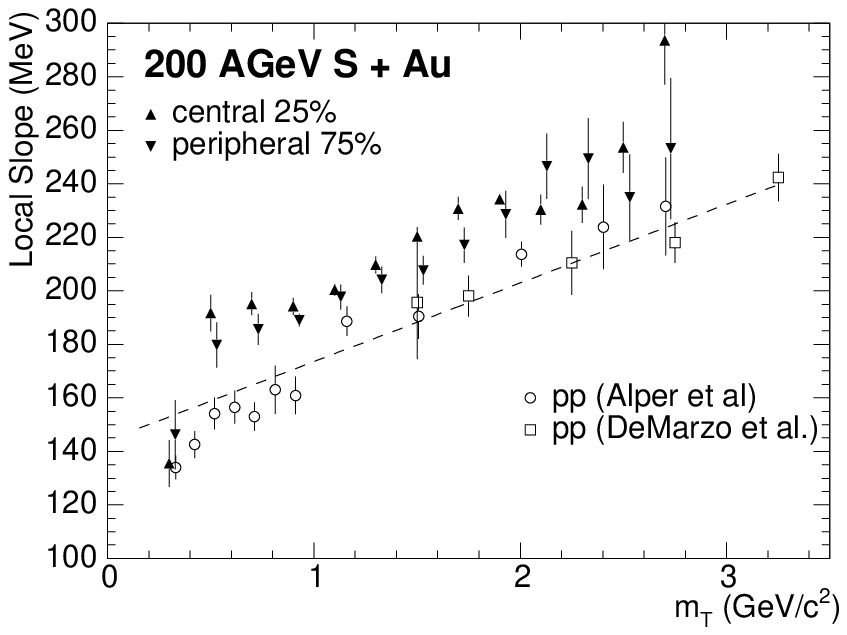}
	\caption[xx]{
		Local slope presentation of central and peripheral S+Au 
		$\pi^0$ data drawn together with minimum bias pp data 
		\cite{ALPER75,NA24-87b}. The dashed line shows the local slopes 
		calculated according to Eq.\,\ref{eqn:pt-local-power} from the fit of 
		Eq.\,\ref{eqn:hagedorn} to the pp data.}
	\protect\label{fig:local-slopes}
  \end{figure}
}

\newcommand{\FigVIII}{
  \begin{figure*}[tb]
\includegraphics{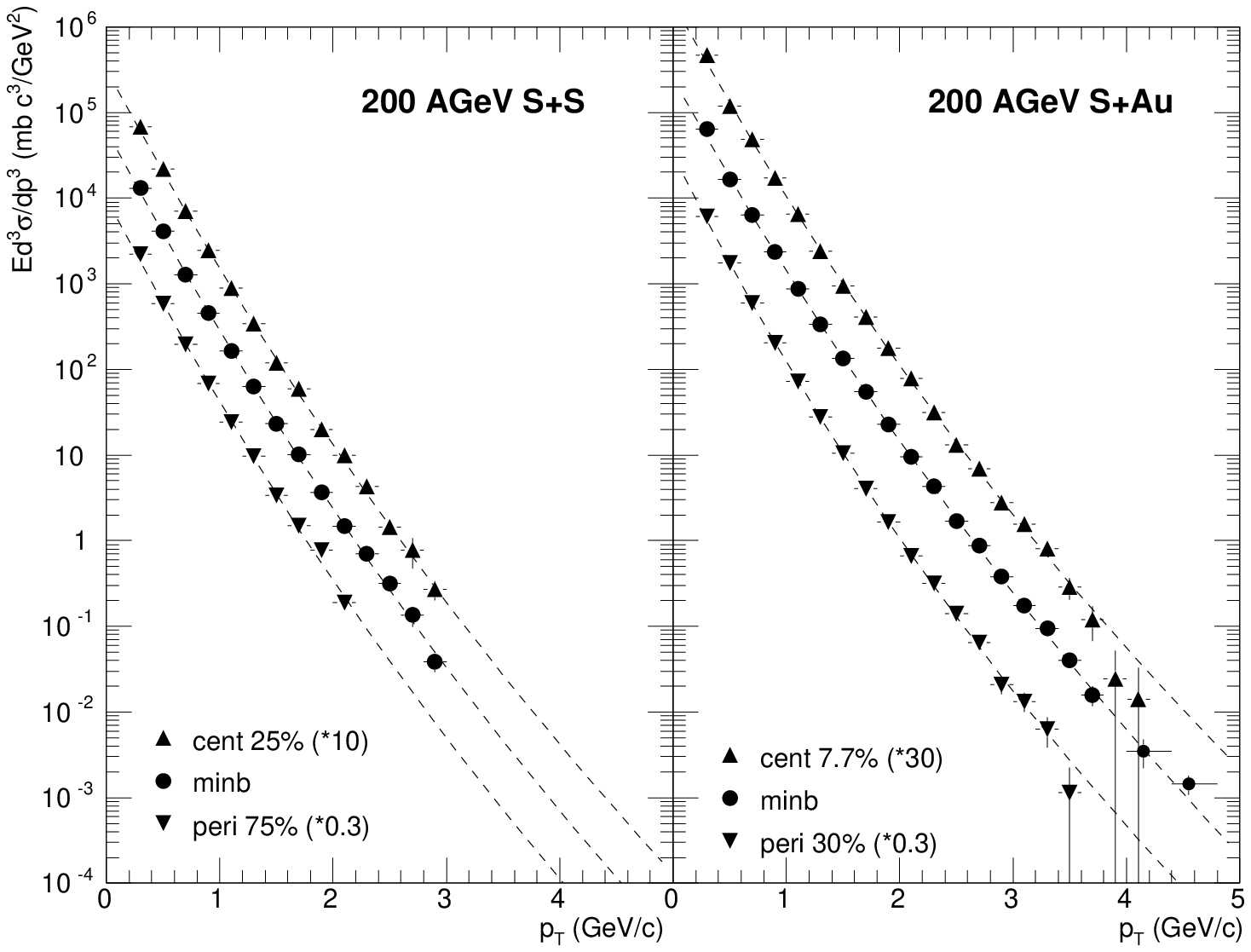}
  \caption{
    Fits to 200 $A\cdot$GeV S\,+\,S and S\,+\,Au data 
    using Eq.\,\ref{eqn:hagedorn} with parameters given in 
    \TabRef{tab:hagedorn}.  
  }
  \protect\label{fig:hagedorn}
  \end{figure*}
  \FigDef{fig:hagedorn}
}

\newcommand{\FigIX}{
  \FigDef{fig:x-ratios}
  \begin{figure}[htb]
  \includegraphics{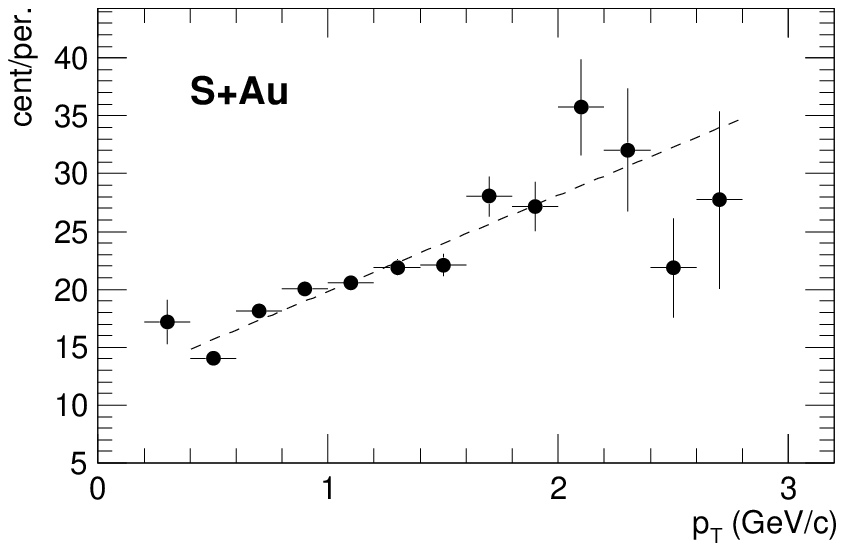}
  \caption{
    Ratio of $\pi^{0}$ multiplicities from central 
    (7.7\%) and peripheral (30\%) S\,+\,Au reactions
    as a function of $p_T$.
  }
  \protect\label{fig:x-ratios}
  \end{figure}
}

\newcommand{\FigX}{
  \FigDef{fig:mean-pt}
  \begin{figure}[htb]
   \includegraphics{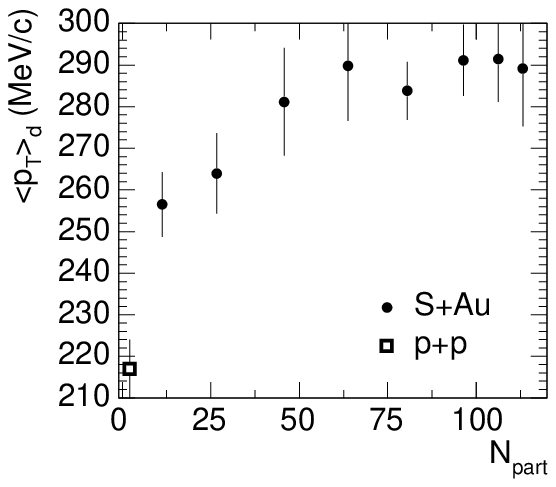}
   \caption{
    Truncated mean transverse momentum $\langle p_T 
    \rangle_{d}$ of $\pi^0$ mesons as defined by 
    Eq.\,\ref{eqn:mean-pt} plotted as a function of the average number of 
    participants $N_{part}$. The data points
    correspond to the 8 $E_{T}$ based centrality selections listed in
    \TabRef{tab:classes}.
    The open square shows $\langle p_T 
    \rangle_{d}$ for pp collisions 
	extracted from the fit with Eq.\,\ref{eqn:hagedorn}. A cut 
	parameter $d = 0.4 \, \mathrm{GeV}/c$ was used.
   }
   \protect\label{fig:mean-pt}
  \end{figure}
}

\newcommand{\FigXI}{
  \FigDef{fig:ratios2}
  \begin{figure}[b]
    \includegraphics{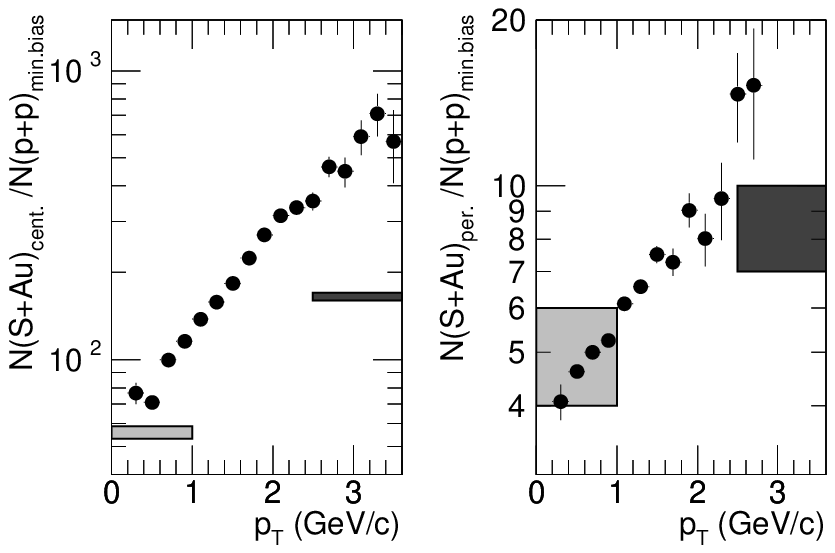}
    \caption[xx]{
      Ratio of central (left) and peripheral (right) 200 
      $A\cdot$GeV S\,+\,Au to p\,+\,p minimum bias $\pi^0$ 
	  multiplicities as a function of $p_T$.  
	  The pp data are scaled to the 
      $\sqrt{s}$ of the present data.
	  Indicated as shaded areas are the ratio of the number of 
	  participants at low $p_{T}$ and the ratio of the number of binary 
	  collisions at high $p_{T}$.
    }
    \protect\label{fig:ratios2}
  \end{figure}
}

\newcommand{\FigXII}{
  \FigDef{fig:cronin}
  \begin{figure}[t]
    \includegraphics{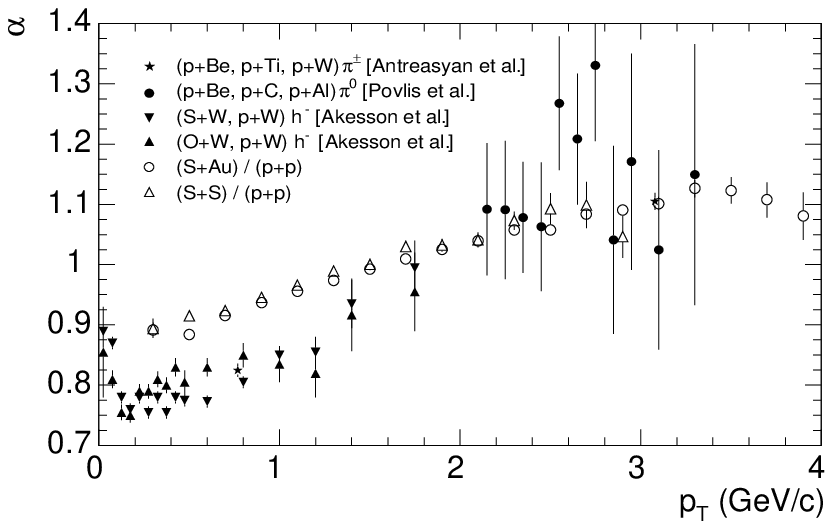}
    \caption[xx]{
      Transverse momentum dependence of $\alpha(p_T)$ from 
      expression \ref{eqn:cronin} for the present $\pi^0$ data from 
      minimum bias S\,+\,S and S\,+\,Au at 200 $A\cdot$GeV 
	  compared with data 
      from other reaction systems \cite{ANTREASYAN79,POVLIS83,na34-90c}.
  }
  \protect\label{fig:cronin}
  \end{figure}
}

\newcommand{\FigXIII}{
  \FigDef{fig:data-string}
  \begin{figure}[b]
    \includegraphics{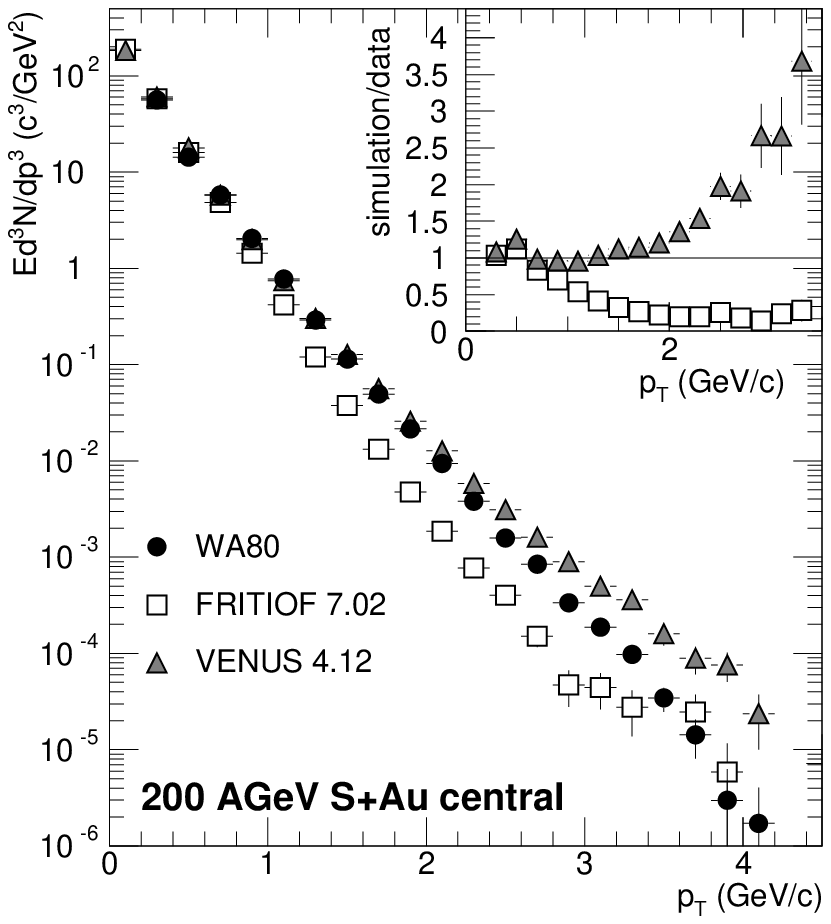}
    \caption[xx]{
      Comparison of the $\pi^{0}$ invariant yields per event 
      of central (7.7\%) 200 GeV S\,+\,Au data with results from 
      string models VENUS \cite{WERNER93a} and FRITIOF \cite{ANDERSSON93}
      both filtered with the experimental acceptance. The inset shows the 
      ratios of the model predictions to the experimental data.
    }
    \protect\label{fig:data-string}
  \end{figure}
}

\newcommand{\FigXIV}{
  \FigDef{fig:schnecki}
  \begin{figure}[bt]
    \includegraphics{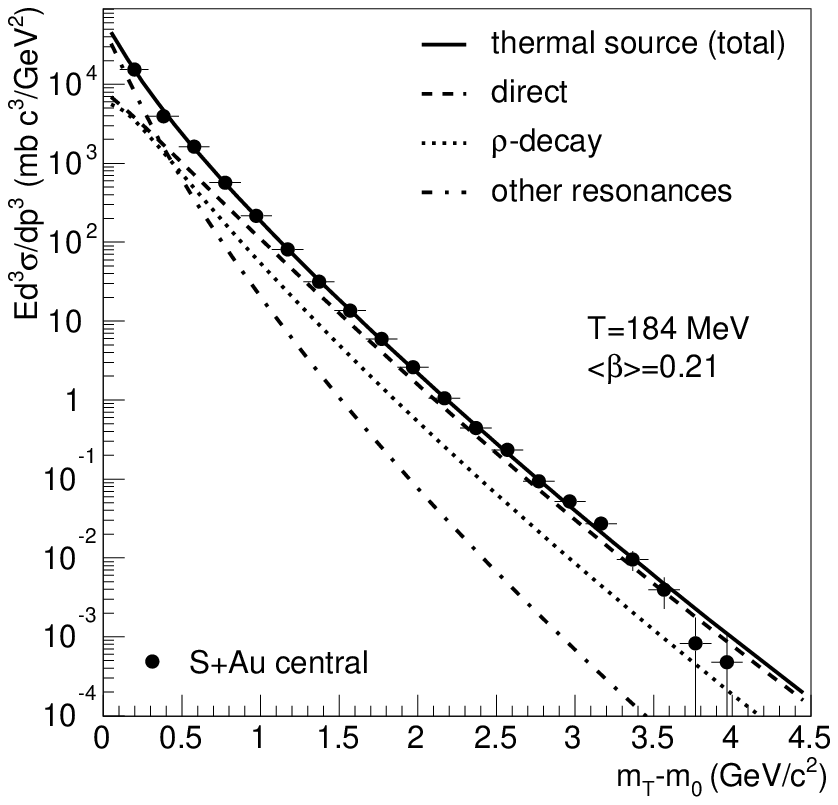}
    \caption[xx]{
      Fits to the $\pi^{0}$ $m_{T}$ distribution of
      central (7.7\,\% of $\sigma_{\mathrm mb}$) 200 GeV S\,+\,Au
      collisions within a thermodynamic model including 
      resonance decays \cite{wiedemann96}.
    }
    \protect\label{fig:schnecki}
  \end{figure}
}

\newcommand{\FigXV}{
  \FigDef{fig:contour}
  \begin{figure}[tb]
    \includegraphics{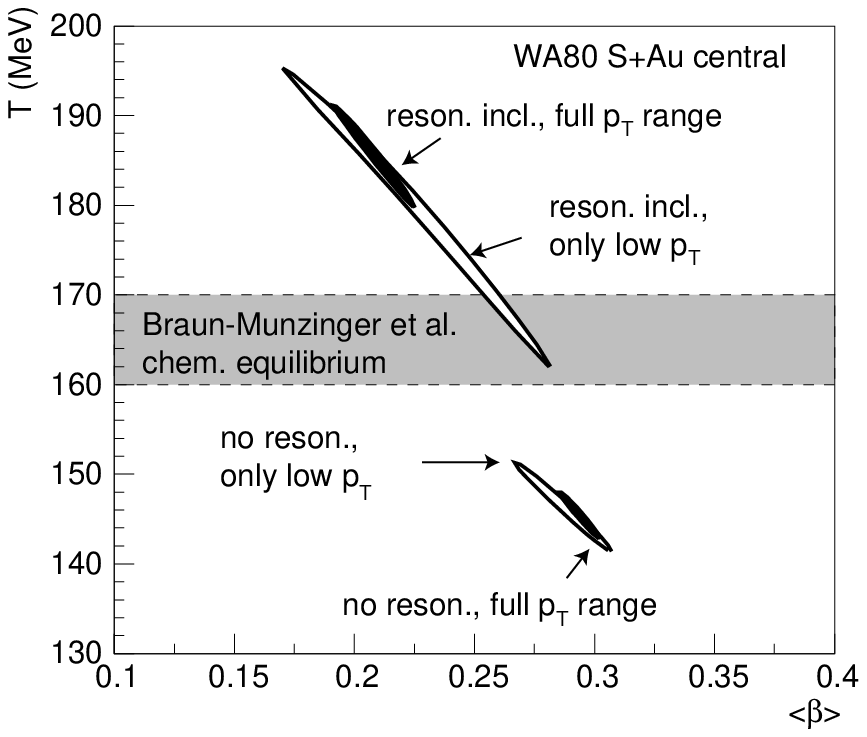}
    \caption[xx]{
      Confidence regions (1 standard deviation) for the fit parameters 
      of a thermodynamic model including 
      resonance decays \cite{wiedemann96} in
      central (7.7\,\% of $\sigma_{\mathrm mb}$) 200 GeV S\,+\,Au
      collisions. The horizontal band indicates the chemical temperature 
      values extracted from particle ratios \cite{braunmunzinger}.
    }
    \protect\label{fig:contour}
  \end{figure}
}

\newcommand{\FigXVI}{
  \FigDef{fig:compfit}
  \begin{figure}[tb]
    \includegraphics{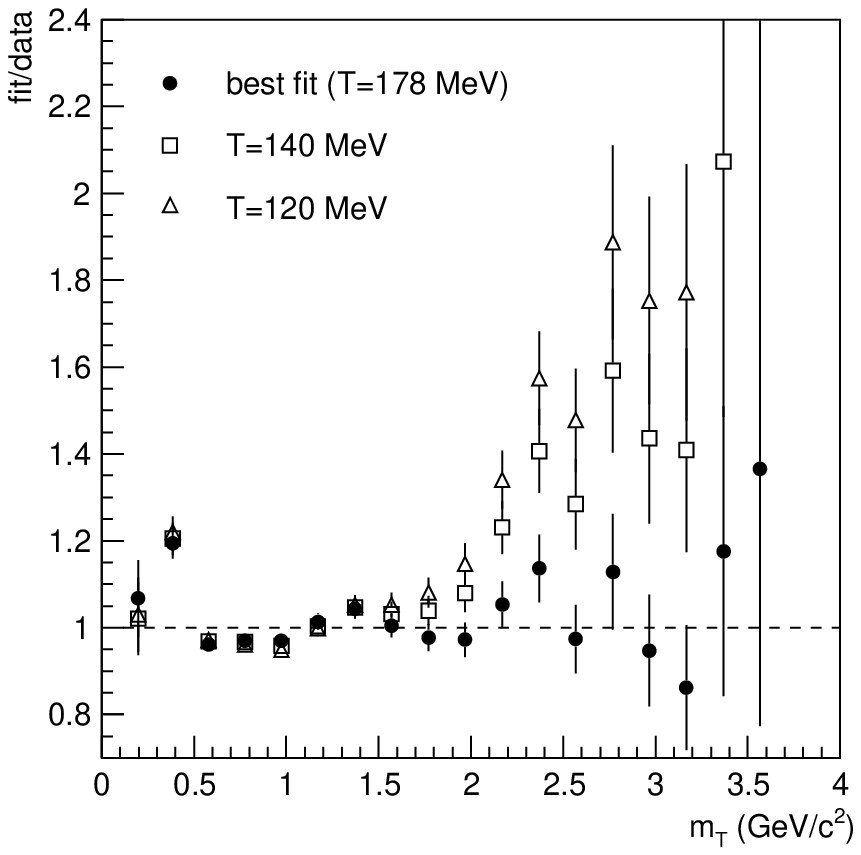}
    \caption[xx]{
      Ratio of fits 
      of a thermodynamic model including 
      resonance decays \cite{wiedemann96} to data for
      central (7.7\,\% of $\sigma_{\mathrm mb}$) 200 GeV S\,+\,Au
      collisions for different temperatures. 
    }
    \protect\label{fig:compfit}
  \end{figure}
}

\begin{document}
\title{Transverse momentum distributions of neutral pions 
from nuclear collisions at 200 $A$GeV}

\subtitle{WA80 Collaboration}

%
%
%

\author{
     R.\,Albrecht\inst{1}
\and V.\,Antonenko\inst{2}
\and T.C.\,Awes\inst{3}
\and C.\,Barlag\inst{4}
\and M.A.\,Bloomer\inst{5}
\and C.\,Blume\inst{4}
\and D.\,Bock\inst{4}
\and R.\,Bock\inst{1}
\and E.M.\,Bohne\inst{4}
\and D.\,Bucher\inst{4}
\and A.\,Claussen\inst{4}
\and G.\,Clewing\inst{4}
\and A.\,Eklund\inst{6}
\and S.\,Fokin\inst{2}
\and A.\,Franz\inst{3}
\and S.\,Garpman\inst{6}
\and F.\,Geurts\inst{7}
\and R.\,Glasow\inst{4}
\and H.\AA.\,Gustafsson\inst{6}
\and H.H.\,Gutbrod\inst{1}
\thanks{now at SUBATECH, Ecole des Mines, Nantes, France}
\and G.\,H\"{o}lker\inst{4}
\and J.\,Idh\inst{6}
\and M.\,Ippolitov\inst{2}
\and P.\,Jacobs\inst{5}
\and R.\,Kamermans\inst{7}
\and K.-H.\,Kampert\inst{4}
\thanks{now at University of Karlsruhe (TH), D-76021 Karlsruhe}
\and K.\,Karadjev\inst{2}
\and B.W.\,Kolb\inst{1}
\and A.\,Lebedev\inst{2}
\and H.\,L\"{o}hner\inst{8}
\and I.\,Lund\inst{8}
\and V.\,Manko\inst{2}
\and S.\,Nikolaev\inst{2}
\and F.E.\,Obenshain\inst{3}
\and A.\,Oskarsson\inst{6}
\and I.\,Otterlund\inst{6}
\and T.\,Peitzmann\inst{4}
\and F.\,Plasil\inst{3}
\and A.M.\,Poskanzer\inst{5}
\and M.\,Purschke\inst{1}
\and H.G.\,Ritter\inst{5}
\and S.\,Saini\inst{3}
\and R.\,Santo\inst{4}
\and H.R.\,Schmidt\inst{1}
\and K.\,S\"{o}derstrom\inst{6}
\and S.P.\,S{\o}rensen\inst{3}\inst{,9}
\and P.\,Stankus\inst{3}
\and K.\,Steffens\inst{4}
\and P.\,Steinhaeuser\inst{1}
\and E.\,Stenlund\inst{6}
\and D.\,St\"{u}ken\inst{4}
\and C.\,Twenh\"{o}fel\inst{7}
\and A.\,Vinogradov\inst{2}
\and G.R.\,Young\inst{3}
}

\institute{
     Gesellschaft f\"{u}r Schwerionenforschung, D-64220 Darmstadt, Germany
\and Russian Research Center "Kurchatov Institute", Moscow 123182, Russia
\and Oak Ridge National Laboratory, Oak Ridge, Tennessee 37831, USA
\and University of M\"{u}nster, D-48149 M\"{u}nster, Germany
\and Lawrence Berkeley Laboratory, Berkeley, California 94720, USA
\and University of Lund, S-22362 Lund, Sweden
\and Universiteit Utrecht/NIKHEF, NL-3508 TA Utrecht, The Netherlands
\and KVI, University of Groningen, NL-9747 AA Groningen, The Netherlands
\and University of Tennessee, Knoxville, Tennessee 37996, USA 
}

\authorrunning{Albrecht et al.}

%
\date{Accepted for publication in Eur. Phys. J. C}
%
\abstract{
New results on transverse mass spectra of neutral 
pions measured at central rapidity 
are presented for impact parameter selected 
200\,$A\cdot$GeV S\,+\,S and S\,+\,Au collisions.  The 
distributions cover more than 8 orders of magnitude in cross section 
over the range $0.3 \, \mathrm{GeV}/c \le 
p_{T} \le 4.0 \, \mathrm{GeV}/c$.  Detailed comparisons to results 
from pp collisions are made.  The spectra from 
all systems show a clear power-law like shape with similar 
curvature. Collisions 
of S\,+\,Au exhibit a larger mean transverse momentum than pp increasing 
with centrality. Predictions of string models and by 
hydrodynamic approaches including collective expansion and decays 
of short lived resonances are compared to the data and the 
implications are discussed.
} 
\maketitle
\section{Introduction}

Heavy ion reactions at sufficiently high energies are able to 
create a strongly compressed and highly excited reaction zone 
which may provide the conditions for the transition to a 
deconfined state of nuclear matter, the Quark-Gluon Plasma.  
Direct information on this state and the initial conditions are 
provided by probes which decouple early in the reaction, like 
photons and dileptons.  Hadrons which interact strongly decouple 
much later and their transverse momentum spectrum therefore gives 
information about the condition of the system at freeze out.  By 
combining the information from photons or dileptons and from 
hadrons and comparing with data from proton-proton interactions 
it is hoped that the development of the heavy-ion system from the 
initial state of high density and temperature through freeze out 
can be reconstructed.  In early pp scattering experiments, it has 
been found that $p_T$ spectra of pions in the range up to about 2 
GeV/$c$ exhibit a nearly exponential shape with a slope constant of 
about 160 MeV$/c$.  This is at first sight surprising since most of 
the pions are not produced directly but emerge from heavier 
resonances with different decay energies.  Furthermore, 
exponential spectra are usually observed in systems, where 
thermodynamical equilibrium is reached through a sizable number 
of rescatterings.  In a proton-proton 
collision this is hardly conceivable.  Using the bootstrap 
hypothesis and particular assumptions on the particle mass 
spectrum, Hagedorn was able to describe the exponential shape of 
the spectra and the limiting value of the slopes 
\cite{HAGEDORN65}.

Proton-proton experiments at the CERN ISR extended measurements to $p_T$ 
values beyond 2 GeV/$c$ \cite{ALPER75,BUESSER73,BUESSER76} and 
subsequently revealed distinct deviations from an exponential shape 
approaching a more powerlaw-like behavior.  This was interpreted as the 
onset of ``hard'' parton scattering \cite{FAESSLER84} and can be 
treated to some extent by perturbative QCD \cite{geist90}.

Data on hadron-nucleus collisions display further characteristic 
differences when compared to proton-proton collisions. These are often 
parametrized by a factor $A^{\alpha(p_T)}$ connecting p+p and 
p+$A$ cross sections.  At low transverse momenta, $\alpha(p_T)$ 
turns out to be about 0.7, which is close to the value of 2/3 expected from 
a geometrical picture of hadron production from an opaque target 
nucleus.  For larger transverse momenta, $\alpha(p_T)$ increases, 
which is interpreted as the onset of parton scattering extending 
over the whole nuclear volume.  In this picture, $\alpha(p_T)$ 
should finally approach the value 1.  The surprising observation 
of $\alpha>1$ in the experimental spectra at high $p_T \ga 3$ 
GeV/$c$ (Cronin-effect \cite{CRONIN75}) was then interpreted as 
the effect of multiple parton scattering \cite{krzywicki79,LEV83}.

All of the features of proton-proton and hadron-nucleus 
scattering described above are expected to be present in 
energetic collisions between heavy nuclei.  In addition, due to 
the nuclear matter being highly excited and compressed over a 
large volume, new specific thermodynamic or hydrodynamic features 
are expected to become apparent.

Theoretically, both the description of nucleus-nucleus 
collisions as a superposition of hadron-hadron collisions 
\cite{sorge89,ANDERSSON93,WERNER93a} as well as the assumption of 
a hot hydrodynamically expanding system of hadrons or partons 
\cite{LEE89,schnedermann93a}, originally proposed by Landau 
\cite{landau53b} are being pursued.

In both approaches, the approximate exponential shape of the final 
spectra at low $p_T$ results from a convolution or superposition of 
various basic processes and sequential decays, and 
cannot easily be associated with a ''temperature''.  At high $p_T$, 
on the other hand, it is expected that the spectra converge to the 
''hard scattering'' shapes observed in proton-proton scattering and 
successfully described by perturbative QCD.

In order to distinguish between the theoretical approaches and to 
isolate possible new features of heavy-ion collisions, it is 
mandatory to investigate the transverse momentum spectra of 
identified particles over a large range in $p_{T}$ extending in 
particular to the hard scattering regime.

In general, the invariant differential cross sections of produced 
particles measured from a given reaction system depend not only 
on the transverse momentum $p_T$, but also on the available 
center-of-mass ({\sl cms}) energy $\sqrt{s}$, the rapidity $y$ of 
the particle, and the associated centrality.  
A comparison of $p_T$ spectra from different 
collision systems or different data sets therefore has to 
carefully consider the dependencies on the quantities noted 
above.

\begin{sloppypar}
In \cite{santo94}, preliminary results of neutral pion spectra for 
S-induced reactions have been presented.
In the present paper, the final results on 
high precision neutral pion $p_T$ spectra 
from collisions of 200 $A\cdot$GeV sulphur nuclei ($\sqrt{s} 
= 19.4$~$A$GeV) with sulphur and gold targets 
in the range 0.3 GeV/$c$ $\le p_T \le$ 4.0 GeV/$c$ and $2.1 
\le y \le 2.9$ for various centralities 
after a complete reanalysis are shown. They will
be compared to data from p\,+\,p and p\,+\,A interactions.  The 
production of $\eta$ mesons measured within the same experiment 
over the $p_T$ range of $0.5 \leq p_T \leq 3.5$ GeV/$c$ \cite{wa80-95c} 
and an upper limit on direct photon production \cite{prl:albrecht:phot:96}
have been recently published.
\end{sloppypar}

\FigI

\section{Experiment}

\begin{sloppypar}
The WA80 experiment at the CERN-SPS (\FigRef{fig:wa80-setup}) 
was equipped with a finely segmented electromagnetic spectrometer 
composed of 3798 lead-glass modules with photomultiplier readout.  
The lead-glass was arranged into three independently calibrated 
arrays, of roughly similar size.  Two arrays consisted of TF1 
lead-glass of $4.0 \times 4.0 \times 40$~cm$^3$ (15 $X_{0}$) 
\cite{berger92} deployed as towers to the left and right of the 
beam axis.  The third array, located below the beam axis, was the 
SAPHIR lead-glass detector \cite{BAUMEISTER90} already used in the WA80 
$^{16}$O run period \cite{WA80-90a} and which consisted of SF5 
lead-glass modules of $3.5 \times 3.5 \times 46$~cm$^3$ (18 
$X_{0}$).  The entire photon spectrometer, which was located at a 
distance of 9\,m downstream of the target, provided coverage of 
10\,\% to 50\,\% of full azimuth $\phi$ over the polar range 
$6.2^\circ \leq \vartheta_{\mathrm lab} \leq 13.9^\circ$.  This 
resulted in a rather uniform acceptance in the rapidity range $2.1 \leq 
y \leq 2.9$ for reconstruction of $\pi^0$ and $\eta$ mesons via 
their $\gamma\gamma$ decay mode.
\end{sloppypar}

\begin{sloppypar}
Immediately in front of the photon spectrometer was a 
double-layer charged particle veto (CPV) counter which covered the 
lead-glass region of acceptance.  Each layer of the CPV 
consisted of streamer tubes with charge-sensitive pad readout, 
with pads of dimension similar to the lead-glass modules 
\cite{albrecht89}.
\end{sloppypar}

The minimum bias (mb) trigger requires a valid signal of the beam 
counters and a minimum amount of transverse energy $E_T \ga 
5$\,GeV, detected by the Mid-Rapidity Calorimeter {\sc mirac} 
\cite{AWES89b}.  Data have been taken with the 200 $A$GeV 
sulfur beam on targets of Au (250~mg/cm$^2$) and S (205 and 
510~mg/cm$^2$).  For the present analysis $7.9$ million S+Au and 
$1.4$ million S+S minimum bias events were accumulated, 
allowing precise reconstruction of the 
$\pi^{0}$ \, $p_T$ distributions over a wide range of $0.3 \leq 
p_{T} \leq 4.3$ GeV/$c$ for S\,+\,Au reactions.  The minimum bias 
cross section has been calculated from the number of beam and 
minimum bias triggers and the target thicknesses, and was 
corrected for small contributions of background effects 
(typically a few percent).  As a result, we obtain $\sigma_{\mathrm mb} 
= 1450$\,mb and 3600\,mb for S\,+\,S and S\,+\,Au reactions, 
respectively.  The S\,+\,Au value is approximately 20\,\% larger than 
for the 1987 WA80 data \cite{WA80-91d}, due to the more biased 
trigger setting used then.
These absolute cross sections have an overall systematic error of 10\%.

\section{Pion Reconstruction and Efficiency}

For the extraction of the neutral pion yield via the 
$\gamma\gamma$ decay branch, all showers in the lead-glass have 
been considered.  While additional criteria like a lateral 
dispersion cut \cite{berger92} or the use of the CPV can improve 
the quality of photon identification, these cuts are of little 
importance for the analysis of neutral pions which are uniquely 
identified via their invariant mass peaks.

\begin{sloppypar}
Hits in the detector are combined to pairs to provide 
distributions of pair mass vs.\ pair transverse momentum for all 
possible combinations.  These distribution are obtained both for 
real events ($R(m_{\mathrm inv},p_{T})$) and for so-called mixed 
events ($M(m_{\mathrm inv},p_{T})$), where a hit from one event 
is combined with a hit from another event with similar 
multiplicity.  $M(m_{\mathrm inv},p_{T})$ provides a good 
description of the combinatorial background. It is subtracted 
from $R(m_{\mathrm inv},p_{T})$ to obtain the mass distribution 
of neutral pions.
\end{sloppypar}

\TabDef{tab:classes}
\begin{table}[h]
\centering
\fbox{\begin{tabular}{|c||r@{\,}c@{}r|r|c|r|}
	\hline
	class & \multicolumn{3}{c|}{$E_{T}$ (GeV)} &
	$\sigma / \sigma_{\mathrm{mb}}$  & cell 
	  & no.\ of \\
	 & \multicolumn{3}{c|}{} &
	  & 
	 occupancy &  events \\
	\hline \hline
	1 &  &$\le$&20.0 & 30.0~\% & 0.17~\% & $2.37 \cdot 10^{6}$ \\
	\hline
	2 & 20.0&$-$&42.0 & 20.1~\% & 0.43~\% & $942  \cdot 10^{3}$ \\
	\hline
	3 & 42.0&$-$&66.3 & 15.2~\% & 0.82~\% & $461  \cdot 10^{3}$ \\
	\hline
	4 & 66.3&$-$&86.0 &  9.5~\% & 1.24~\% & $539  \cdot 10^{3}$ \\
	\hline
	5 & 86.0&$-$&107.5 & 10.5~\% & 1.55~\% & $1.48 \cdot 10^{6}$ \\
	\hline
	6 & 107.5&$-$&121.5 &  7.0~\% & 1.90~\% & $1.02 \cdot 10^{6}$ \\
	\hline
	7 & 121.5&$-$&133.5 &  4.9~\% & 2.10~\% & $717  \cdot 10^{3}$ \\
	\hline
	8 &  &  $>$ & 133.5 &  2.8~\% & 2.40~\% & $407 \cdot  10^{3}$ \\
	\hline
	\end{tabular}}
	\caption{Centrality classes, as selected by the amount of 
	transverse energy observed in {\sc mirac}, and cell
	occupancies for $^{32}$S~+~Au collisions.}
	\label{tab:classes}
\end{table}

In central reactions of $^{32}$S~+~Au the average particle 
occupancy in the detector reaches values of $2.4\,\% / 
\mathrm{module}$ (see \TabRef{tab:classes}).  This leads to a 
finite probability that showers in the detector overlap and 
influence each other.  Such an overlap may result in the 
following modifications of the measured signals:

\begin{enumerate}
	\item  The signal of one shower is ``absorbed'' by another 
	one -- this happens mostly to low energy showers. In such a 
	case, the photon is lost and if it originated from a 
	$\pi^{0}$-decay, the $\pi^{0}$ parent particle itself cannot
	be reconstructed and is lost.

	\item  The signal of one shower is influenced by that of a nearby 
	shower. In this case, the measured energy and/or position may 
	be changed and, thereby, the momentum of the photon may be changed.
	If it originated from a neutral pion, depending on how important
	these changes are, this may either result in
	\begin{enumerate}
		\item  changes of the apparent invariant mass such that the 
		pion cannot be reconstructed, or
		\item  changes that still allow to identify a pion by its 
		proper invariant mass, but at a different momentum\label{pi0mod}.
	\end{enumerate}
	In addition, case \ref{pi0mod} leads to a modification of the shape
	of the invariant mass distribution of $\pi^{0}$s, which has to be
	taken into account in the extraction of the peak content.
\end{enumerate}

These effects lead to a detector efficiency for pion 
reconstruction which depends on particle density. To study this 
detection efficiency, the GEANT \cite{geant} simulation package  
has been used to create artificial signals for the lead glass 
modules corresponding to neutral pions hitting the detector. In 
total, $\approx 10^{6} \, \pi^{0}$s have been simulated with uniformly 
distributed transverse momenta. 

\FigIII

The signals were then analyzed with the same analysis programs as 
used for real data.  Analyzing them as single particles, allows 
to study detector effects and the analysis programs.  To obtain 
a $\pi^{0}$ efficiency, however, these simulated photon pairs have to be 
studied in a realistic particle density environment.  This has 
been done by superimposing them onto real measured events.  
Effects of detector noise and digitization of the photomultiplier 
signals are implemented.  
Most importantly, the simulation provides the 
means to extract the probability for a pion at given input 
transverse momentum $p_{T}^{(0)}$ to be measured with 
$p_{T}^{(1)}$.  These probabilities measured as a function of 
$p_{T}^{(0)}$ can be understood as a matrix which transforms real 
physical distributions into measured ones.  The {\em efficiency 
matrix} contains both effects of detection efficiency as 
described above, as well as the geometrical acceptance. 
The acceptance can be easily separated and is shown in 
\FigRef{fig:acceptance} as a function of the transverse momentum.
The efficiency matrix is 
calculated for each event class (i.e.\ given particle 
density).  The representation as a matrix has the advantage 
that the actual shape of the momentum distributions does not need 
to be known beforehand.

\FigIV

The efficiency matrices are used to correct the measured 
distributions in the following way:
\begin{enumerate}
\item  A reasonable $\pi^{0}$ distribution is chosen as initial ansatz.
\item  The distribution is transformed by applying the efficiency 
matrix.
\item  The resulting distribution ($B$) is compared to the
actual measured distribution ($A$).
\item  The input distribution is modified by multiplying it
with the ratio $A/B$.
\item  Steps 1-4 are iterated until $A$ and $B$ agree.
\end{enumerate}

The result of this procedure is shown in
\FigRef{fig:efficiency}, where the $\pi^{0}$ reconstruction 
efficiency\footnote{We refer now to the ``pure'' efficiency without 
any effect from the geometrical acceptance.} 
is plotted for the most peripheral and most central 
S\,+\,Au reactions, respectively.  It can be seen that for very 
peripheral reactions, the correction is relatively small, 
particularly for $p_{T} \ge 1$\,GeV/$c$.  However, for the most 
central reactions, i.e.\ for the highest particle densities, the 
correction becomes much more important.  The general behavior can 
be understood from the previous discussion. The drop in the 
$\pi^{0}$ reconstruction efficiency at low $p_{T}$ is largely 
caused by the increasing probability with decreasing energy 
to misidentify or lose a 
low energy photon in the radial tail of a higher energy shower, 
such that the (mostly low $p_{T}$) neutral pion cannot be 
reconstructed and is lost for further analysis.  At large 
transverse momenta, the reconstruction efficiency exceeds the 
value 1.  This is an effect of limited transverse momentum 
resolution on steeply falling spectra; effectively the observed 
distribution is broadened over the original one by the apparatus, 
thus giving rise to $\varepsilon_{\pi^{0}} > 1$ (item 2b in the 
list above). 

This is illustrated in \FigRef{fig:overlap}. Here, 
pions have been simulated with transverse momenta in a narrow range
($0.8 \, \mathrm{GeV}/c \leq p_{T} \leq 1.0 \, \mathrm{GeV}/c$)
The resulting $p_{T}$ 
distributions after the overlap procedure discussed above are shown
as dark grey histograms. One 
can see that already in peripheral collisions (upper right) there is 
some broadening of the distribution, the effect is however much 
stronger in central collisions (lower right). Here a significant 
fraction of the pions will be measured with higher transverse 
momentum. For the steeply falling spectra indicated as line histograms 
this may result in a larger apparent number of pions in the higher 
$p_{T}$ bins and thus lead to an ``efficiency'' $> 1$. An equivalent 
broadening effect can be seen in the experimental data e.g. in the 
invariant mass spectra displayed on the left hand side of 
\FigRef{fig:overlap}. Overlap effects lead to an additional high mass tail 
of the pion peak in central collisions. 

\FigII

The systematic errors on the measured transverse momentum spectra are 
dominated by the following contributions\footnote{The estimates given 
here apply to central collisions where the errors are largest.}:
\begin{itemize}
	\item An uncertainty in the absolute calibration of the momentum 
	scale of 1\%. This may be translated into an uncertainty in the 
	yields\footnote{This error contribution 
	is much smaller in the ratios of 
	photon or pion spectra as e.g. used for the direct photon analysis
	\cite{prl:albrecht:phot:96}.} 
	of a few \% at low $p_{T}$ rising to $\approx 13 \%$ at 
	$p_{T} = 3.5 \, \mathrm{GeV}/c$.
	
	\item Uncertainties in the $\pi^{0}$ extraction which include the 
	error in the determination of the invariant mass peak content and 
	the propagation of the energy resolution through the acceptance and 
	efficiency corrections. This leads to an error of 16\% at 
	$p_{T} = 0.5 \, \mathrm{GeV}/c$ which decreases to below 3\% at 
	high $p_{T}$.
	
\end{itemize}
These two major contributions to the systematic errors and their 
quadratic sum are shown in \FigRef{fig:fehler}.

\FigV

\section{Results}
\subsection{Nuclear Reactions}

\FigVI

Efficiency corrected $\pi^0$ invariant cross sections measured in 
the rapidity range $2.1 \leq y \leq 2.9$ are presented in 
\FigRef{fig:pt-ss-sau} for S\,+\,S and S\,+\,Au reactions.  The 
data are shown for central, peripheral and minimum bias trigger conditions.  
The central event class of 
7.7\,\% of $\sigma_{\mathrm mb}$ in 
S\,+\,Au collisions corresponds to complete overlap of the S 
nucleus with the Au target, with an average of $\approx 110$ participating 
nucleons, to be compared to an average of 
$\approx 11$ participating 
nucleons for the 30\,\%  
most peripheral events.
From a first inspection it is seen that the slopes appear systematically 
flatter for central collisions than for peripheral 
ones.  This might indicate a higher temperature in more violent 
central collisions.  Although the slope parameters cannot be 
identified \emph{a priori} with a temperature in a thermodynamic sense, 
we shall often use the label ``temperature'' in the 
following.\footnote{All slope parameters will be given in units MeV.}  
Superimposed on the data are exponentials $d\sigma/dm_T^{2} 
\propto \exp(-m_T/T)$ fitted to the region 0.8 GeV$/c^{2}$ $\le m_T 
\le 2.0$ GeV$/c^{2}$. The fitted slope parameters are indicated.  In 
this parameterization of the intermediate $m_T$ data, the 
flattening of the distributions with increasing centrality 
results in a higher apparent temperature.  The high precision and 
the large $m_T$ range of the present $\pi^0$ data allows 
to recognize, however, that the spectral shapes are only very 
poorly described by an exponential curve.  Instead, a concave 
behavior reminiscent of pp data at similar or higher $\sqrt{s}$ 
is observed over the whole $m_T$ 
range.  Its properties are more systematically analyzed by 
calculating the local slope, defined as %
\begin{equation}
T_{\mathrm loc}^{-1}=-\frac{d}{dm_T} \left[
     \log \left( E\frac{d^3\sigma}{dp^3} \right) \right] \: .
\end{equation}
\NOTE{omit local slope at $p_{T} = 0.3$?}
The results are plotted in \FigRef{fig:local-slopes} for the 
S\,+\,Au data.  Here, each individual value of $T_{\mathrm loc}$ 
has been determined locally from three adjacent data points of 
\FigRef{fig:pt-ss-sau}.  Such an analysis puts very high 
demands on the statistical and systematic uncertainties in order 
to keep point-to-point fluctuations at an acceptable level.  
Shown together with the heavy ion data are pp minimum bias data 
which will be discussed in section\,\ref{sec:pp-pa}.

As expected, the observed local slopes are not constant,
as would be the case for a purely exponential distribution, but change 
continuously over the measured $m_T$ range both for peripheral 
and central collisions.  This again indicates that fits with one 
or two exponentials, which have often been presented elsewhere 
(e.g.\ Ref.~\cite{NA35-94a}), can only provide a crude
approximation valid over a restricted $m_T$ range.  
\FigRef{fig:local-slopes} also demonstrates that the slope 
parameters in the intermediate $m_T$ range are systematically 
higher for central compared to peripheral data.

\medskip

In order to systematically compare data from different systems 
and/or taken under different trigger conditions it is often more 
convenient to apply a description in terms of a functional form.  
According to the above discussion, such a parameterization must 
allow for $p_T$ dependent variations in the curvature of the 
spectra.  At high $p_T$, where perturbative QCD becomes 
applicable, the spectra are expected to attain a power-law 
behavior, as observed in numerous high energy pp measurements (see 
for example Ref.~\cite{UA1-96}).  The heavy-ion data of this 
experiment seem to follow that trend even into the intermediate 
$p_{T}$ range.  Therefore, a parameterization originally inspired 
by QCD \cite{HAGEDORN83} and successfully applied already to
pp data \cite{UA1-96} has been chosen to fit to the spectra:
\begin{equation}
\label{eqn:hagedorn}
E\frac{d^3\sigma}{dp^3} = C \left( \frac{p_0}{p_T+p_0} \right)^n
\end{equation}
with $C$, $p_0$ and $n$ taken as free parameters.  A link to the 
more familiar exponential slope parameter $T$ is obtained from 
the derivative of Eq.\,\ref{eqn:hagedorn} according to
\begin{equation}
\label{eqn:pt-local-power}
T_{\mathrm power-law} =
- \frac{f(p_T)}{\frac{\partial f(p_T)}{\partial p_T}} =
\frac{p_0}{n} + \frac{1}{n} \cdot p_T \: .
\end{equation}

\FigVII

\FigVIII

\TabDef{tab:hagedorn}
\begin{table*}[bt]
	\caption[xx]{
		Parameters obtained by fitting Eq.\,\ref{eqn:hagedorn} to the 
		invariant cross section of neutral pions from different reaction
		systems and for different classes of centralities. The parameter
		$T_1$ and $T_2$ are local slopes according to Eq\,\ref{eqn:pt-local-power}.
		The pp data taken at $\sqrt{s}=23.5$\,GeV have been scaled to
		$\sqrt{s}=19.4$\,GeV before fitting according to the recipe given in 
		\cite{beier78}.}
	\label{tab:hagedorn}
\newcommand{\RU}{\rule[-2mm]{0mm}{7mm}}
\newcommand{\MR}{\multicolumn{2}{|c|}{\RU}}
\newcommand{\MC}[1]{\multicolumn{2}{|c|}{#1}}
\newcommand{\CL}{\cline{3-16}}
\newcommand{\HL}{\hline}
\newcommand{\PM}{|r@{$\,\pm$\,}r}
\newcommand{\NE}[1]{\multicolumn{2}{|c|}{#1}}

\centering
\mbox{
\small
\begin{tabular}{|r@{\,+\,}l|c|r@{\,\%\,}l \PM \PM \PM \PM \PM |r|}
\hline
\MC{\rule[-2mm]{0mm}{7mm}} & &
 \MC{Fraction} &
  \MC{$p_0$} &
   \MC{$n$} &
    \MC{$p_0/n$} & 
     \MC{$T_1$ at 1 GeV} &
      \MC{$T_2$ at 3 GeV} &
        $\chi^2_{\mathrm red}$
         \\
\MC{} & & 
 \MC{of $\sigma_{\mathrm mb}$} & 
  \MC{(GeV$/c$)} &
   \MC{} &
    \MC{(MeV$/c$)} &
     \MC{(MeV)} &
       \MC{(MeV)} &
	\\
\HL
\MR      &centr.& 7.7 && 5.6 & 0.5 & 32.0 & 2.0 & 174.5 & 2.6 & 205.7 & 1.0 & 268.1 & 3.2 & 2.2 \\ \CL
\MR      &centr.&  25 && 5.7 & 0.3 & 32.8 & 1.2 & 174.3 & 1.5 & 204.7 & 0.5 & 265.7 & 1.9 & 3.2 \\ \CL
\RU S&Au &      & 100 && 6.5 & 0.3 & 36.4 & 1.4 & 178.6 & 1.4 & 206.1 & 1.8 & 261.0 & 1.8 & 2.2 \\ \CL
\MR      &per.  &  75 && 5.3 & 0.4 & 31.6 & 2.0 & 166.7 & 2.4 & 198.3 & 0.7 & 261.5 & 3.5 & 0.6 \\ \CL
\MR      &per.  &  30 && 4.9 & 0.5 & 30.2 & 2.3 & 163.0 & 2.8 & 196.1 & 0.8 & 262.3 & 4.3 & 0.6 \\ \HL

\MR      &centr.&  25 && 5.3 & 1.4 & 31.7 & 6.8 & 166.2 & 7.5 & 197.7 & 2.4 & 260.8 & 9.8 & 1.0 \\ \CL
\RU S&S  &      & 100 && 5.6 & 1.4 & 33.5 & 6.6 & 166.8 & 6.5 & 196.6 & 8.7 & 256.4 & 8.7 & 0.7 \\ \CL
\MR      &per.  &  75 && 6.2 & 4.1 & 36.2 &20.5 & 170.2 &11.5 & 197.9 & 2.6 & 253.2 &19.2 & 1.2 \\ \HL

\RU p&p  &      & 100 && 4.9 & 0.3 & 34   & 1   & 146   & 2   & \NE{175}    & \NE{234}    & 4.45\\ \HL

\end{tabular}
}
\end{table*}

Thus, $p_0/n$ characterizes the slope of the transverse momentum 
spectra in the limit $p_T \rightarrow 0$, while $1/n$ 
characterizes its gradient along $p_T$, i.e.\ the strength of the 
concave curvature.  The analytical form of 
Eq.\,\ref{eqn:pt-local-power} can in that way be compared 
directly to the numerical values of the local slopes presented in 
\FigRef{fig:local-slopes}.
As stressed by Hagedorn \cite{HAGEDORN83}, 
and immediately recognized from Eq.\,\ref{eqn:pt-local-power}, 
$p_0$ and $n$ are highly correlated. We have therefore chosen to 
express Eq.\,\ref{eqn:hagedorn} as a function of the local slopes 
$T_{1}$ and $T_{2}$ at two fixed values of $p_{T}$ (1\,GeV$/c$ and 
3\,GeV$/c$) -- these are less correlated and therefore better suited as 
fit parameters.

Fits to the data using Eq.\,\ref{eqn:hagedorn} are presented in 
\FigRef{fig:hagedorn}.  The extracted parameters are compiled 
in \TabRef{tab:hagedorn} together with the results of 
analogous fits to p\,+\,p data \cite{ALPER75}.
All slope parameters are observed to increase smoothly 
with the centrality 
of the reaction, e.g. $T_{1}$ increases from approximately 196\,MeV in 
peripheral collisions to approximately 206\,MeV in central collisions.

%
\FigIX

Another way to investigate the differences between peripheral and 
central data in a model independent way is to study the ratio of 
the transverse momentum distributions as a function of $p_T$.  
\FigRef{fig:x-ratios} shows that with increasing $p_T$, the 
pion multiplicity in central data is enhanced relative to 
that of peripheral data.  Measured over the dynamic range of the 
S\,+\,Au data, a variation by a factor of $\ga 2$ is observed.  
Similar findings in the cross section ratios have been reported 
earlier by us for central and peripheral O\,+\,Au reactions at 
200 GeV \cite{WA80-90a}.

The mean transverse momentum $\langle p_T \rangle$ is an 
observable containing important information about the reaction 
dynamics of the system and is often used to provide a convenient 
means of comparison among different experiments and model 
calculations.  The evaluation of $\langle p_T \rangle$ is usually 
achieved analytically by using parameterizations of the 
experimental distributions (see below).  Unless the cross section 
is measured down to very low $p_T$ and the functional form 
describes the data well, this method may lead to systematic 
errors due to the dominance of the low $p_{T}$ cross section.  
If only the variation of $\langle p_T \rangle$ as a function of 
another observable, like for example the centrality, is of 
importance, it is more advantageous to calculate instead the 
truncated mean transverse momentum above a certain threshold 
directly from the experimental data.  This can be done by the 
relation:
\begin{equation}
	\langle p_{T} \rangle_d = \left( \left.
	\int_{d}^{\infty} p_{T} \frac{dN}{dp_{T}} dp_{T} \right/
	\int_{d}^{\infty} \frac{dN}{dp_{T}} dp_{T} \right) - d
	\label{eqn:mean-pt}
\end{equation}
Here, $d = 0.4$ GeV/$c$ is chosen according to the lowest $p_{T}$ 
data point where systematic uncertainties imposed by acceptance 
and efficiency corrections are still well under control.  For a 
purely exponential invariant cross section, 
i.e.\ $d\sigma/dp_{T}^{2} \propto \exp(-p_{T}/T)$, and for $d=0$, 
one immediately finds $\langle p_T \rangle_{d=0} = 2T$.  For finite 
values of $d$, however, this simple relation does not hold and 
$\langle p_{T} \rangle_d$ decreases as a function of $d$.

Most interesting in connection with the thermodynamical picture 
of the compressed and excited heavy ion system is the variation 
of $\langle p_T \rangle$ with $E_T$ or the multiplicity 
$dN_{ch}/d\eta$ of produced particles.  
We therefore have studied $\langle p_T \rangle_{d}$ of identified 
$\pi^0$ mesons as a function of the 
transverse energy. It has been demonstrated earlier \cite{WA80-91d} 
that $E_{T}$ is proportional to the total number of participating 
nucleons $N_{part}$. For S\,+\,Au reactions $N_{part}$ has been 
calculated for the different centrality cuts by Glauber calculations. 
In \FigRef{fig:mean-pt} the values of $\langle p_T \rangle_{d}$ as a 
function of $N_{part}$ are shown. 
A rise in $\langle p_{T} \rangle_d$ with 
increasing $N_{part}$ is observed, which 
again reflects the increase with centrality of the slope 
parameters of the exponential fits in \FigRef{fig:pt-ss-sau}, 
or of the local slopes shown in \FigRef{fig:local-slopes}. The 
values seem to saturate at $\approx 290 \, \mathrm{MeV/c}$ for 
medium-central to central collisions. These relatively small values 
may be compared with those obtained from a complete integration 
of the fits with Eq.\,\ref{eqn:hagedorn}. For central S\,+\,Au 
collisions one obtains $\langle p_T \rangle \approx 380 \, \mathrm{MeV/c}$ -- 
these values, however, have a large uncertainty due to the low 
$p_{T}$ extrapolation. 

\FigX

\subsection{Comparison to pp and p\mbox{\boldmath$A$} data}
\label{sec:pp-pa}
The transverse momentum spectra presented for S\,+\,Au reactions 
unambiguously demonstrate distinct variations as a function of 
event centrality.  In order to learn, how these 
characteristic changes can be attributed to differences in the 
size of the interacting system, it is of interest to compare with 
data from pp and pA collisions in a consistent way.  
Unfortunately, a rigorous comparison cannot be made free of 
assumptions, since the relevant data from different experiments, 
in general, do not correspond to the same $\sqrt{s}$ and rapidity 
range.  The large body of experimental data collected for pp 
reactions, however, may be used to extract appropriate scaling 
laws of transverse momentum spectra as a function of $\sqrt{s}$ 
and rapidity $y$ and thus allow to adopt these data to the 
corresponding values of our experiment.  In this work, we have 
employed the empirical concept which is presented in 
Ref.~\cite{schmidt93b} together with data from 
Refs.~\cite{ALPER75,BUESSER73,BUESSER76,beier78,ANTREASYAN79,NA24-87b}. 
The minimum bias pp data of Ref.~\cite{ALPER75,NA24-87b}, taken 
at $\sqrt{s} = 23.5$~GeV and at slightly more forward 
rapidities, were thus scaled appropriately to the conditions of 
the WA80 experiment. 
The systematic uncertainty introduced by 
this procedure is included in the error bars shown. 

A first comparison of the $\pi^{0}$ spectra from pp collisions 
with that of S-induced reactions was shown in 
\FigRef{fig:local-slopes}.  The slope parameters 
$T_{\mathrm loc}$ smoothly grow from approx.\ 150~MeV at $m_T 
\approx 0.3$~GeV/$c^{2}$ to almost 250~MeV at $m_T \ga 
3.0$~GeV/$c^{2}$, demonstrating convincingly the strong concave shape 
of the transverse momentum distributions. The local slopes both in 
p\,+\,p and heavy ion data increase in a similar way when going from 
low $p_{T}$ to high $p_{T}$, thus indicating that the curvature is 
very similar in both cases.\footnote{As discussed above, small 
differences in the curvature can be seen from the fits of 
Eq.\,\protect\ref{eqn:hagedorn}.} The main difference is in the 
systematically higher values of the local slopes in heavy ion 
reactions.
The slopes of the $p_T$ spectra in pp and nucleus-nucleus collisions 
differ over the whole $p_T$ range which indicates the occurence of 
strong nuclear effects in particle production. 


The variations seen in the local slopes are reflected in the 
truncated mean transverse momentum (\FigRef{fig:mean-pt}). Here,
$\langle p_{T} \rangle_d$ has also been calculated for pp data from 
a parameterization. The value of 217~MeV/$c$ is considerably below 
those for the S\,+\,Au data, even for the most peripheral collisions. 
Apparently, the number of participating nucleons even in the 
peripheral collisions ($\approx 11$) is too large to treat them as 
identical to pp collisions.

\FigXI

\FigXII

The ratio of multiplicities is shown in \FigRef{fig:ratios2} 
as a function of $p_T$ for the same set of data.  In this 
presentation, we find an increase of the 
(S\,+\,Au)$_{\mathrm{cent}}$/pp ratio 
with $p_T$ by almost a factor of 10 over the 
dynamic range of the experiment and for 
(S\,+\,Au)$_{\mathrm{per}}$/pp 
an increase by a factor of $\approx 4$. 
We would like to point 
out that this large increase of the 
cross section ratio at a given $p_T$ may be seen as a consequence
of the differences of the local slope at lower values of $p_T$  
(see \FigRef{fig:local-slopes}). 
The ``high $p_T$ enhancement'' of 
the cross sections therefore appears not to be an independent feature 
at high $p_T$ but may rather be considered as
the result of the different slopes over 
the complete $p_T$ range.

For comparison \FigRef{fig:ratios2} includes estimates of the ratios 
of the number of participants (light grey area at low $p_{T}$) and of 
the number of binary collisions (dark grey area at high $p_{T}$). The 
pion production seems to scale approximately with the number of 
participants at low $p_{T}$, but increases much more strongly at high 
$p_{T}$ in central S\,+\,Au collisions compared to pp than expected 
from the number of binary collisions.

A similar variation of the cross section ratios was already 
observed by Cronin {\sl et al.} \cite{CRONIN75} in a comparison 
of p\,+\,p and p\,+\,A data.  The systematic behavior could be 
successfully parameterized by the phenomenological expression
\begin{equation}
\label{eqn:cronin}
E\frac{d^3\sigma}{dp^3} ({\mathrm{p+A}}) =
A^{\alpha (p_T)} \: E \frac{d^3\sigma}{dp^3} ({\mathrm{p + p}})
\end{equation}
where, apart from the expected $\sqrt{s}$ dependence, the 
parameter $\alpha(p_T)$ was found to depend only weakly on rapidity 
\cite{chaney79} but strongly on $p_T$.  At lower $p_T$, $\alpha$ 
turned out to be approximately 0.7 which in  
Eq.\,\ref{eqn:cronin} gives a factor close to $A^{2/3}$ as expected 
for an opaque target nucleus.  For higher transverse momenta, 
$\alpha$ was found to increase with $p_T$ indicating a growing 
participation of the nuclear volume, approaching $\alpha = 1$ in 
the limit of full participation of the target nucleus.  The 
surprising observation of $\alpha > 1$ for the highest $p_T$ 
values, called `anomalous nuclear enhancement', was subsequently 
interpreted as an effect of multiple scattering of the incident partons 
\cite{LEV83}.

\begin{sloppypar}
For $A$\,+\,$B$ collisions (where $A$ and $B$ denote heavy 
nuclei), the straightforward extension of Eq.\,\ref{eqn:cronin} 
writes
\begin{equation}
E\frac{d^3\sigma}{dp^3} ({\mathrm{A+B}}) =
(A \cdot B)^{\alpha (p_T)} \: E\frac{d^3\sigma}{dp^3} ({\mathrm{p + p}})
\label{eqn:cronin2}
\end{equation}
and has successfully been applied to experimental data in 
Ref.~\cite{na34-90c}. \FigRef{fig:cronin} shows a compilation 
of $\alpha(p_T)$ obtained from the present data together with 
other data sets. The investigated projectiles, targets, and 
detected particle species are indicated together with their 
references. 
One should note that 
the statistical accuracy of the present data even at high $p_{T}$ is 
comparable to the data of \cite{ANTREASYAN79} 
and is superior to that of the other data sets.
\end{sloppypar}

It appears that Eq.\,\ref{eqn:cronin2} provides an almost 
universal parameterization for all systems with $\alpha > 1$ 
at high $p_T$. The differences of the exponent of the present 
data compared to NA34 at low $p_T$ \cite{na34-90c} may be caused by 
the more backward rapidity region of the NA34 experiment. However, 
the data for S\,+\,S and S\,+\,Au data do also disagree with the data 
of \cite{ANTREASYAN79} at $p_{T} = 0.77 \, \mathrm{GeV}/c$ which are 
measured in a similar rapidity region. It should be noted 
that the different trigger biases are another source of systematic 
errors in such a comparison of minimum bias cross sections. It may 
also be relevant that the present data are compared to a compilation 
of pp data, while in \cite{ANTREASYAN79} one particular data set 
measured within the same experiment has been used. Different 
normalizations of the pp data could therefore contribute to a 
difference in $\alpha$.

\section{Discussion}

The slope and the curvature of the transverse momentum spectra, 
studied as a function of projectile energy, rapidity, and 
multiplicity, carry important information about the reaction 
dynamics.  All spectra of produced particles emerging from pp, 
p$A$, and $AB$ collisions are known to exhibit in a reasonable 
approximation an exponential shape which might be regarded as 
indication for a thermal emission mechanism.  In fact, for a 
thermalized system the particle $p_{T}$ spectra reflect the 
temperature of the system at the time of decoupling.  This 
idealized picture is altered, if there is collective motion in 
the system, such as hydrodynamic flow.  In such a case, the 
transverse velocities of particles are to be considered a sum of 
a random thermal component and an ordered collective expansion 
velocity.  If one or both components change between different 
reaction systems or impact parameter regions, the 
transverse momentum distributions are expected to be affected. 
The picture of 
thermal emission has, despite its simplicity, proven very 
successful.  The assumptions made in that interpretation might, 
however, be questioned particularly in pp systems because of the 
small volume available for particle production.  Surprisingly, 
the same concept has recently even been applied to hadron 
production in $e^+ e^-$ collisions at LEP energies and was able 
to reproduce many of the observed particle ratios 
\cite{becattini96}. One may take the success of these models as a 
hint for the validity of statistical concepts, which however do not 
necessarily imply a thermal origin.

With the growing amount of data and their improved precision it 
has been well established that $p_T$-spectra of all systems, when 
measured with sufficient accuracy and over a large $p_T$ range, 
display a curved structure which cannot be fit by a simple 
exponential.  Therefore, fits often have been made with either a 
single exponential in a restricted $p_T$ region and deviations 
have then been termed ``low $p_T$ enhancement'' or ``high $p_T$ 
enhancement'' respectively, or a superposition of two or more 
exponentials has been applied over a larger range of $p_{T}$.  
Without a physical picture, the former approach is rather 
misleading since it arbitrarily defines a ``normal'' behavior and 
deviations from that, and the latter one easily leads to an 
inflation of fit parameters. 

On the other hand, there are in fact indications for different 
mechanisms at high and low $p_T$ which have to be treated 
separately and should be included appropriately in a complete 
description of the full $p_T$ spectrum.  For $p_T \ga 2$\,GeV/$c$ 
spectra from pp collisions have successfully been described by 
perturbative QCD calculations \cite{UA1-96}.  The ``nuclear 
effects'' observed in p$A$ reactions have then been discussed in 
terms of multiple parton scattering.  A generalization and 
extension of these calculations for AB collisions is provided 
by the kinetic parton cascade model \cite{GEIGER92a,GEIGER92b}, 
which treats the reaction -- for processes of large momentum 
transfer -- on the basis of interacting quarks and gluons.  Since 
this model employs perturbative QCD, it can only be applied to 
large $p_T$ processes.  For the much more abundant low $p_T$ 
processes, presently only phenomenological models can be applied.

Several codes are available, which are based on 
phenomenological string models 
\cite{sorge89,ANDERSSON93,WERNER93a} and use experimental data 
from elementary collision processes as input.  Within these 
models, after appropriately accounting for the nuclear geometry, a 
heavy ion collision is treated as a superposition of individual 
nucleon-nucleon collisions.  Obviously, such an approach neglects 
any collective effect which might be present in an extended 
strongly interacting system.  Comparison with various data shows 
that such a straightforward extrapolation from pp collisions 
provides a reasonable description, particularly for the 
intermediate $p_T$ range.  This indicates that minimum bias heavy 
ion collisions in this region do not exhibit dramatic effects 
extending beyond the straightforward extrapolation from pp 
collisions.  Deviations from standard string model simulations 
are observed particularly for central nuclear collisions both at 
low and high $p_T$.  Various attempts have thus been made to 
improve the description of the data by including additional 
mechanisms into the string models.  These extensions include hard 
scattering processes \cite{ANDERSSON93}, rescattering of the 
produced particles in the target nucleus and among themselves 
\cite{WERNER93a}, production of quark clusters \cite{WERNER95a}, 
or color rope formation \cite{sorge95}.

\FigXIII

As an example, VENUS 4.12 \cite{WERNER93a} and FRITIOF 7.02 \cite{ANDERSSON93} 
calculations are shown in \FigRef{fig:data-string} together 
with the $\pi^0$ spectrum from central S+Au collisions.  The 
simulated events have been filtered with the experimental 
acceptance and no normalization to the cross section of the data 
is applied.  The description of the spectra is reasonably good at 
$p_{T} \leq 1 \, \mathrm{GeV}/c$. At higher $p_T$, FRITIOF falls 
below the data and is about a factor of 5 too low at 
$p_{T} \geq 2 \, \mathrm{GeV}/c$. The agreement between data and VENUS 
is better, which may be attributed to the rescattering of produced 
particles implemented in this model. VENUS however also starts to 
deviate at higher $p_T$; it is more than a factor of 2 too high at 
$p_{T} \geq 3 \, \mathrm{GeV}/c$.
If rescattering in the nuclear medium is included, the description at 
high $p_T$ is improved somewhat, at the expense, however, of 
additional parameters in the model. Still the description is not 
perfect.

A completely different approach starts from nuclear hydrodynamics 
and treats the heavy ion collision as an expanding 
thermodynamical system characterized by a temperature, chemical 
potential, and expansion velocity 
\cite{LEE89,schnedermann93a,schnedermann,wiedemann96}.  Especially for 
central collisions of heavy nuclei thermal equilibrium seems 
plausible, at least locally, due to the frequent rescattering of 
produced particles in the extended medium.  This picture is 
supported by the large rapidity shifts of produced particles 
observed in central compared to peripheral collisions \cite{WA80-91d,WA80-92d}.  
The extraction of freeze-out temperatures from measured pion $p_T$ spectra 
within this model is complicated for several practical and 
conceptual reasons.  First, it has been recognized that a 
significant fraction of the observed pions originate from decays 
of short-lived heavier resonances.  This leads to characteristic 
deviations from a single exponential and, particularly at low 
$p_{T}$, to a more curved spectrum, as observed experimentally 
\cite{na27-91,graessler78}.  By including all contributing 
resonances, the authors of \cite{sollfrank91} were able 
to describe the experimental $p_T$ spectra of pions from pp 
systems.  The success of this approach, however, does not 
necessarily prove that thermal equilibrium is in fact reached 
in such systems.

\FigXIV

The same procedure has been applied to heavy ion systems using a 
computer program provided by the authors of \cite{wiedemann96}. Here, 
however, in addition to the most important resonance decays, transverse 
flow has also been included in the calculations. This 
leads to a ``blue shift'' of the apparent temperatures relative 
to the freeze-out temperatures of the system.  By fixing the 
baryonic chemical potential to $\mu_{B} = 200\, \mathrm{MeV}$ 
and varying the temperature and 
expansion velocity, good fits to the experimental data have been 
obtained.  An example of such a calculation 
applied to the present data is shown in \FigRef{fig:schnecki}.

While in earlier analyses
of $p_T$ spectra these models were not sensitive enough to fix 
simultaneously the values of the temperature and the radial 
expansion velocity without ambiguities, the precision of the 
present data allows to further restrict the parameter values. The 
spectrum is best described with a temperature
$T = 184 \, \mathrm{MeV}$ and an average transverse flow 
velocity of $\langle \beta \rangle = 0.21$.
The fits are not very sensitive to the baryonic chemical potential 
because baryon resonances contribute mostly at very low $p_{T}$ where 
the experimental coverage vanishes. This can be seen from
\FigRef{fig:schnecki}, where the total yield is broken down into 
the direct thermal pions (dashed line), the contribution from $\rho$ 
decays (dotted line) and all 
other resonances (dash-dotted line).

\FigXV

\FigRef{fig:contour} shows 1-sigma-confidence regions in the 
($T$,$\left\langle \beta \right\rangle$) parameter plane for these fits. 
Calculations have been performed under the following conditions:
\begin{enumerate}
	\item  Including all resonances using
	\begin{enumerate}
			\item  the full experimental $m_{T}$-range (solid area in the 
			upper part of \FigRef{fig:contour}) and
		
			\item  only data for $m_{T} < 1.6 \, \mathrm{MeV}/c^2$ 
			(outline in the upper part of \FigRef{fig:contour})
		\end{enumerate}
		and

	\item  including only the direct pions using the same $m_{T}$ 
	regions (a) and (b) as above (solid area and outline in the lower 
	part of \FigRef{fig:contour}).
\end{enumerate}

A fit to the entire $m_{T}$ range with the full model constrains the 
temperature to the range $T = 180 - 190 \, \mathrm{MeV}$, a fit to 
the restricted $m_{T}$ range yields essentially the same optimal 
parameters ($T = 178 \, \mathrm{MeV}$), but relaxes the 
constraints -- still the 1-$\sigma$ range of the temperature is above 
160~MeV. The 2-$\sigma$ confidence region is essentially unbound 
towards lower temperatures when using 
only data for $m_{T} < 1.6 \, \mathrm{MeV}/c^2$, one can, however, 
see that such fits for lower temperatures are in contradiction with 
the data at higher $m_{T}$. \FigRef{fig:compfit} shows the ratio of 
fits to the experimental data for the optimum value ($T = 178 \, 
\mathrm{MeV}$) and constrained fits for $T = 140 \, \mathrm{MeV}$ and 
$T = 120 \, \mathrm{MeV}$. It can be seen that the fits for lower 
temperatures overpredict the cross section at higher $m_{T}$. In 
the presence of additional hard scattering contributions the 
experimental data would still provide an upper bound for the hydrodynamical 
contribution. Therefore the full $m_{T}$ range may be used to estimate 
a lower bound of the temperature parameter.

It is interesting to note that fits ignoring the resonance 
contributions would indicate much lower temperatures and also slightly 
higher flow velocities than those of the full calculation.
In this case the temperature and flow velocity are rather close to the 
values given in \cite{na44-97} for S\,+\,S collisions 
($T \approx 140 \, \mathrm{MeV}$, $\langle \beta \rangle = 
0.24-0.28$).\footnote{The S+S data in the present publication yield 
similar results to the S+Au data, their statistical 
uncertainty is, however, much larger so that they have not been considered 
further.} 
Similar numbers are given in \cite{sollfrank97}. One 
should keep in mind, however, that there is no good reason to exclude 
the resonance decays, and that our results for the full calculation 
are not consistent with the values in \cite{na44-97,sollfrank97}.

\FigXVI

While the calculation with resonances is clearly more reasonable than 
the one without, there are still further uncertainties related to the 
model assumptions. 
Especially the choice of one temperature to characterize both  
chemical freeze-out of all resonances and thermal freeze-out of 
the pions is questionable. The chemical freeze-out temperature 
of the $\rho$ meson as a resonance appearing in $\pi\pi$ 
scattering might be closely related to the thermal freeze-out 
temperature of the pions. However, this is most likely not true for 
other particles like e.g. the $\mathrm{K}^{\star}$. It is therefore 
necessary to add information from other measurements, e.g. relative 
particle abundancies or interferometry measurements.

Due to the finite value of the expansion velocity, 
i.e.\ an energy sharing between random thermal and ordered 
collective motion, the resulting temperature 
is now smaller than the slope parameter obtained from the 
exponential fits in \FigRef{fig:pt-ss-sau}. The temperature of 184~MeV 
can be compared to the value of $T = 160-170 \, \mathrm{MeV}$ 
extracted from relative particle abundancies \cite{braunmunzinger}.


\section{Summary}

We have discussed transverse momentum spectra of 
neutral pions produced in various systems.  
The extracted local slopes as well as the power-law fits to the 
pp and AB \, $p_{T}$ spectra unambiguously demonstrate 
concave distributions for all systems and impact parameter 
selections.  
The shapes of the spectra can be well characterized by two parameters, 
e.g. the inverse slope at a given $p_{T}$ ($T = T_{1}$) and the linear 
variation of the slope parameters ($\Delta T = T_{2} - T_{1}$). While 
$\Delta T$ (the ``slope of the local slope'') and thereby the 
curvature are very similar in all systems, 
there is a strong increase 
of the inverse slope parameters and  of $\langle p_{T} \rangle$ in 
going from pp to heavy ion collisions and a continuing, but more 
subtle growth with the size of the 
interaction zone.\footnote{Corrections for the 
influence of a shift of the effective center-of-mass on the transverse 
momentum spectra measured as a function of centrality in asymmetric 
heavy-ion collisions have not been considered here, although they 
might -- within certain thermal models -- account for part of the 
centrality dependence of the slope parameters.}  

The comparison to event generators shows, that they still fail to describe 
all subtle features observed in the data if measured over a large 
dynamic range. The FRITIOF model shows large discrepancies, VENUS 
4.12 with its phenomenological rescattering provides a better, yet still 
unsatisfactory description. A hydrodynamically motivated calculation 
with two fitted parameters agrees very nicely with the data. Within 
such a calculation both transverse flow and decay of 
long-lived resonances appear to be of importance for the 
interpretation of measured transverse momentum spectra. The 
fit applied to the data in the present paper favours comparably large 
freeze-out temperatures and low flow velocities. As such large 
temperatures might require high pion phase space densities, it has to 
be investigated whether they may be interpreted as freeze-out 
temperatures or whether the model used here with the assumption of 
local thermal equilibrium may be not applicable.


The spectra are broadened when going from pp to AB collisions. 
At first sight, ratios of the type 
pA/pp, AB/pp, central/peripheral, etc.\ seem to isolate the 
existence of strong high $p_{T}$ phenomena.  A closer look at the 
individual changes of the spectral shapes, however, allows to 
trace the increasing ratios with increasing $p_{T}$ back to the 
differences in the overall shape of the spectra which necessarily 
leads to large effects in the cross section at large $p_{T}$. The 
difference in the slopes over the full $m_{T}$ range may be taken as 
a hint for non-trivial nuclear effects observed in these collisions.
The observed 
variations may, in the framework of thermodynamics, be taken as 
an indication of higher temperature and/or a possible stronger 
collective expansion of the system in heavy ion collisions compared 
to pp.  Likewise, in the picture of 
string models, it may suggest a stronger amount of rescattering 
of produced particles. Another possible conclusion would be to call 
for different mechanisms, i.e. string formation and decay as 
sufficient for pp collisions, but local thermodynamic equilibrium and 
hydrodynamic expansion in heavy ion collisions.

%
Common to all 
models, even the most conservative ones, is the need to either 
incorporate large amounts of initial- and final-state scattering, 
or to assume high temperature and pressure build up.  Therefore, 
they imply a dense and strongly interacting system, i.e.\ 
conditions which are favorable to eventually reach thermal 
equilibrium.

\begin{acknowledgement}
This work was supported jointly by the German BMBF and DFG, the U.S. DOE, 
the Swedish NFR, the Humboldt Foundation, the International Science
Foundation under Contract N8Y000, the INTAS under Contract
INTAS-93-2773, ORISE, and the Dutch FOM. ORNL is managed 
by Lockheed Martin Energy Systems under contract DE-AC05-84OR21400 with the 
U.S. Department of Energy. We also like to thank the accelerator 
divisions of CERN for their excellent work.
\end{acknowledgement}
%

\begin{thebibliography}{10}

\bibitem{HAGEDORN65}
R.~Hagedorn, Suppl. Nuovo Cimento {\bf 3} (1965) 147--186.

\bibitem{ALPER75}
B.~Alper et~al., Nucl. Phys. {\bf B100} (1975) 237--290.

\bibitem{BUESSER73}
F.W. B{\"u}sser and et~al, Phys. Lett. {\bf 46B} (1973) 471--476.

\bibitem{BUESSER76}
F.W. B{\"u}sser et~al., Nucl. Phys. {\bf B106} (1976) 1--30.

\bibitem{FAESSLER84}
M.~A. Faessler, Phys. Rep. {\bf 115} (1984) 1--91.

\bibitem{geist90}
W.M. Geist et al., Phys. Rep.
  {\bf 197} (1990) 263--374.

\bibitem{CRONIN75}
J.W. Cronin et~al., Phys. Rev. {\bf D11} (1975) 3105--3123.

\bibitem{krzywicki79}
A.~Krzywicki et al., Phys. Lett. {\bf B85}
  (1979) 407--416.

\bibitem{LEV83}
M.~Lev and B.~Petersson, Z. Phys. {\bf C21} (1983) 155--161.

\bibitem{sorge89}
H.~Sorge, H.~St{\"o}cker, and W.~Greiner, Nucl. Phys. {\bf A498} (1989)
  567c--576c.

\bibitem{ANDERSSON93}
B.~Andersson, G.~Gustafson, and H.~Pi, Z. Phys. {\bf C57} (1993) 485--494.

\bibitem{WERNER93a}
K.~Werner, Phys. Rep. {\bf 232} (1993) 87--299.

\bibitem{LEE89}
K.S. Lee and U.~Heinz, Z. Phys. {\bf C43} (1989) 425--429.

\bibitem{schnedermann93a}
E.~Schnedermann et al., Phys. Rev. {\bf C48} (1993)
  2462--2475.

\bibitem{landau53b}
L.~Landau, {Proc. Acad. Sci., USSR (Phys. Series)} {\bf 17} (1953) 51.

\bibitem{santo94}
R.~Santo et~al., Nucl. Phys. A {\bf 566} (1994) 61c--68c

\bibitem{wa80-95c}
{R. Albrecht et al., WA80 Collaboration}, Phys. Lett. {\bf B361} (1995) 14--20.

\bibitem{prl:albrecht:phot:96}
R.~Albrecht et~al., Phys. Rev. Lett. {\bf 76} (1996) 3506

\bibitem{berger92}
F.~Berger et al., Nucl. Instr. Meth. {\bf A321} (1992) 152--164.

\bibitem{BAUMEISTER90}
H.~Baumeister et al., Nucl.
  Instr. Meth. {\bf A292} (1990) 81--96.

\bibitem{WA80-90a}
{R. Albrecht et al., WA80-Collaboration}, Z. Phys. {\bf C47} (1990) 367--375.

\bibitem{albrecht89}
R.~Albrecht et al., Nucl. Instr. Meth. {\bf A276} (1989) 131--139.

\bibitem{AWES89b}
T.~C. Awes et~al., Nucl. Instr. Meth. {\bf A279} (1989) 479--502.

\bibitem{WA80-91d}
{R. Albrecht et al., WA80-Collaboration}, Phys. Rev. {\bf C44} (1991)
  2736--2752.

\bibitem{geant}
{R.~Brun et~al.}
\newblock {GEANT3, CERN Data Handling Division}.
\newblock {DD/EE/84-1}, 1987.

\bibitem{NA35-94a}
{J. B{\"a}chler et al., NA35 Collaboration}, Phys. Rev. Lett. {\bf 72} (1994)
  1419--1422.

\bibitem{UA1-96}
{G. Bocquet et al., UA1-Collaboration}, Phys. Lett. {\bf B366} (1996) 434.

\bibitem{HAGEDORN83}
R.~Hagedorn,
\newblock CERN-TH. 3684, 1983.

\bibitem{beier78}
E.W. Beier et~al., Phys. Rev. {\bf D18} (1978) 2235--2238.

\bibitem{NA24-87b}
{C. DeMarzo et al., NA24 Collaboration}, Phys. Rev. {\bf D36} (1987) 16--20.

\bibitem{schmidt93b}
H.R. Schmidt and J.~Schukraft, J. Phys. {\bf G19} (1993) 1705--1796.

\bibitem{ANTREASYAN79}
D.~Antreasyan et~al., Phys. Rev. {\bf D19} (1979) 764.

\bibitem{POVLIS83}
J.~Povlis et al., Phys. Rev. Lett. {\bf 51} (1983)
  967--970.

\bibitem{na34-90c}
{T. {\AA}kesson et al., NA34-Collaboration}, Z. Phys. {\bf C46} (1990)
  361--367.

\bibitem{chaney79}
D.~Chaney et~al., Phys. Rev. {\bf D19} (1979) 3210--3221.

\bibitem{becattini96}
F.~Becattini, Z. Phys. {\bf C69} (1996) 485--492.

\bibitem{GEIGER92a}
K.~Geiger, Phys. Rev. {\bf D46} (1992) 4965--4985.

\bibitem{GEIGER92b}
K.~Geiger, Phys. Rev. {\bf D46} (1992) 4986--5005.

\bibitem{WERNER95a}
K.~Werner and J.~Aichelin, Phys. Rev. {\bf C52} (1995) 1584--1603.

\bibitem{sorge95}
H.~Sorge, Phys. Rev. {\bf C52} (1995) 3291--3314.

\bibitem{schnedermann}
E.~Schnedermann, Phys. Rev. {\bf C50} (1994) 1675.

\bibitem{wiedemann96}
U.A.~Wiedemann and U.~Heinz, Phys. Rev. {\bf C56} (1997) 3265-3286.

\bibitem{WA80-92d}
{R. Albrecht et al., WA80-Collaboration}, Z. Phys. {\bf C55} (1992) 539--548.

\bibitem{na27-91}
{M. Aguilar-Benitez et al., NA27-Collaboration}, Z. Phys. {\bf C50} (1991)
  405--426.

\bibitem{graessler78}
H.~Gr{\"a}ssler et~al., Nucl. Phys. {\bf B132} (1978) 1--14.

\bibitem{sollfrank91}
J.~Sollfrank et al., Z. Phys. {\bf C52} (1991) 539.

\bibitem{na44-97}
I.G.~Bearden et al., Phys. Rev. Lett. {\bf 78} (1997) 2080--2083.

\bibitem{sollfrank97}
J.~Sollfrank et al., \emph{Mass number scaling in ultra-relativistic 
nuclear collisions from a hydrodynamical approach}, preprint 
nucl-th/9801023 and BI-TP 97/55.

\bibitem{braunmunzinger}
{P. Braun-Munzinger et al.}, Phys. Lett. {\bf B365} (1996) 1-6.

\end{thebibliography}

%
%
%

\end{document}